**Methodological opportunities for mitigating climate change in complex food systems**

Erik van der Linden[1), a] *, Marcel Meinders[2)], Egbert H. van Nes [3)], Hugo de Vries[4), a]

1) Laboratory of Physics and Physical Chemistry of Foods, Department of Agrotechnology and Food Sciences, Wageningen University, Bornse Weilanden 9, 6708 WG Wageningen, The Netherlands
2) Wageningen Food and Biobased Research, Bornse Weilanden 9, 6708 WG Wageningen, The Netherlands
3) Laboratory of Aquatic Ecology and Water Quality Management, department of Environmental Sciences, Droevendaalsesteeg 3. 6708PB Wageningen, The Netherlands
4) UMR IATE, CIRAD, INRA, Montpellier SupAgro, University Montpellier, 2 Place Viala, 34060 Montpellier, France

* Corresponding author

[a] Shared last author

**Abstract**

Unravelling current complex food systems is relevant for their adjustment and redesign under the current changing climate conditions. Redesign may be necessitated by migration of people and changes of locations of major agri-food production. The redesign should be conducted synchronously with that of systems entangled with the food system, such as the socio-economic and cultural system. For such synchronous redesign a common methodological approach with a common set of methods is required. In the current article we suggest a common set of methods, and discuss how these methods find their basis in vastly different science fields, ranging from soft matter, biology, urban socio-economics, ecology, to machine learning. We address the various ways such methods have been applied in relatively small parts of the food systems and how they can be applied to larger parts of current and future food systems. The set of methods facilitates to identify the level of structuredness and randomness in complex systems. It helps to better predict upcoming transitions in complex systems, according critical points, and sudden instabilities. It facilitates in extracting information from a system, before, during and after the time that one makes an intervention, which in turn will help to decide which interventions are best to maintain or change functions of a complex system. Further progress on addressing complex challenges like mitigating climate change in relation to sustainable food production is foreseen by using the universal concepts and combining analytical matrix features of dominance versus randomness, domain specific encoding of neural networks and identifying dominancy in neural networks using the analytical matrix features of dominance versus randomness.

## 1. Introduction

Food systems may be defined as combinations of materials and/or operations that provide foods, including the food products themselves. Adequately managing food systems is crucial to ensuring global food security. The challenge relates, among other aspects, to identifying and continuously adapting interventions tailored at mitigating climate change, biodiversity loss, wars, political tensions, shortage of water, food harvest insecurity, or its sudden abundance. In earlier papers (van Mil, Foegeding et al. 2014, Perrot, De Vries et al. 2016, de Vries, Donner et al. 2022) a systematic approach for unravelling complex food systems was proposed, yet to an abstract level. Three main aspects were proposed (van Mil, Foegeding et al. 2014): a) determine the specific problem or topic in terms of its conceptual level and (spatial and temporal) scale of observation; b) obtain information about levels and scales at that level and below (in term of the elements and interactions at these scales); c) develop models at the different spatial scales and combining them into one larger model. The approach was formulated in terms of a roadmap of procedures that use an array of methods applied in an integrative multiscale and inter-disciplinary way. In follow-up papers actual models incorporating the roles and behavior of actors have been discussed (Perrot, De Vries et al. 2016, de Vries, Donner et al. 2022).

The current article discusses in more detail the methods to effectively extract information in more detail from a system in order to be better able to manage it. The system examples we use are of a soft matter, biological, city-societal, or ecological nature. We discuss the particular methods applied to one or more of these system examples. We include some recent advances for these examples in using such methods, and illustrate how several methods are already being combined. As such the current article provides a more detailed conceptual framework for addressing points a) to c) than in the previous work by Van Mil et al.

Each of these methods have been separately applied to other complex systems but their combination is argued to provide a better systematic approach to unravelling complex food systems. We address specific examples of food systems in the various sections first, and then discuss them overall in our discussion section.

Complex systems have received increasing attention from the 1980's onwards and are omni-present and are often societally relevant. They include, for example, animals, plants, the brain, immune systems, ecologies, economies, cities, and global production systems (Waldrop 1993). We refer to publications of Anderson introducing the topic from a physics point of view and addressing the fundamental aspects of studying such systems (Anderson 1972, Anderson 1995). The importance of work on complex systems often has a curiosity driven element to it, and an element that is inspired from its usefulness.

We note that all such complex systems are characterized by the fact that they consist of elements that are exhibiting self-organization, i.e. allowing adaptation to changing internal and external conditions, where the organization of the elements is partially responsible to the system properties and its functions.

Complex systems has casted some mystery around it. On the book by Waldrop entitled: Complexity, the emerging science between order and chaos, it is Douglas Hofstadter who remarks about the book: This is a deep tale of <u>science in the making</u> …. (Waldrop 1993). This suggests a new type of science. However, Gershenson (Gershenson 2008) remarks in his book "one of the most debated aspects on complexity is its status of its science". One can add to this, as a consequence, whether complex systems need a new science. Indeed, in regards to enabling investigations and lending from frameworks we already have at our disposal, Bar-Yam remarks that any specific domain has developed tools to address the specific complex systems of that domain, and therefore that "… many of the tools can be adapted for more general use by recognizing their universal applicability" (Bar-Yam 1997). It is such universal applicability of specific concepts that we address in this article. It revolves around the universal notion how a function or property of a system can be described by one or more dominant features, and statistically determined features. We point to an existing universality existing across fields ranging from soft matter, urban socio-economics, plant sciences, animal sciences, ecology, and machine learning (AI).

An important example of a complex system is the food system, as especially articulated during the United Nations Food Systems Summit in 2021 (Braun, Afsana et al. 2021). Both planetary challenges (like climate change and biodiversity loss), as well as societal challenges (political tensions, poverty and growing inequalities) clearly urge for better insights in local and global complex systems, to better predict outcomes of interventions on such systems, facilitating appropriate approaches and actions to transform current systems for a more sustainable future. In the global food system, the availability of raw materials and natural resources is compromised by the different demands of various intertwined systems like the food, water, and energy system, whereas these are also essential to human kind. The need for predictions of intervention outcomes on all these scales is increasing, e.g. for human health, sustainable environment, and global economy. The combination of such global complex systems, while including climate changes, yields an even larger and more overwhelming complexity. One of the main challenges for the future will be how to sustain food production systems amidst the sustaining of other systems, and identify key determinants that facilitate such sustainability. We refer to a recent publication by King and Jones that shed some light on such key determinants (King and Jones 2021). In this perspective, the role of AI in precision agriculture is indicated by for example Khan et al. (Khan, Ray et al. 2021). It is noted that analysis leads one to believe "ambitious conservation efforts and food system transformation are central to an effective post-2020 biodiversity strategy", which should be included in all efforts (Leclère, Obersteiner et al. 2020). Part of the food system transformation may be related to shifting to a plant based diet (Ritchie 2021). In general terms, a change on a global scale and thus a food system transition in this case, can occur when convictions, values and norms that are rooted in institutions are synchronously moving into an average direction of change by small steps (Chapin, Weber et al. 2022),(Scheffer 2026).

Finding an adequate description of a combination of complex entities will be facilitated when applying common approaches and methods for unravelling the separate complex systems. This makes a common set of methods with common concepts important.

Apart from the necessity to understand food systems to optimise and adapt them, the global structure of food systems may very well need to be altered more drastically due to climate changes, most likely due to changes in the location of people around the globe [(Adger, Crépin et al. 2020)] in combination (Scheffer 2025) with according changes in magnitude of food crisis exposure (Strona 2025). From the above it is clear that food security is a multi-faceted problem, including demographic, social-cultural elements, political, environmental and economic elements. The results by Strona allow a more fact based exploration of socio-economic scenario's, while including possible demographic and poverty projections (Strona 2025). Interestingly, a focus on environmental and social sustainability would significantly reduce the number of severe food crises people would be exposed to.

As the term complexity is used throughout the entire article, for all systems, we will start with a quantification of the term complexity and introduce some important concepts alongside. Secondly, as many of the methods we will address are all lent from physics, we choose to then address the complex soft matter systems to introduce these methods and discuss their relevancy for those systems. ubsequently we address the other system classes mentioned, including food systems, and end with a summary and conclusions.

## 2. Quantification of the term complexity.

We apply a practical and theoretically sound measure for the term "complex", as used in the title, and, alongside, the terms "complexity" and "information".

In analyzing any system, complex or simple, one needs to obtain information from the system. Usually, the information is spread over different characteristic scales, and over different mutually interacting physical and virtual entities at these different scales. The multitude of interactions and the multitude of different types of entities result in the complexity of the system itself and of its analysis. Until here, the terms information and complexity are used loosely, but they can be defined in a quantitative and rigid manner.

In general, when something is referred to as complex it refers to the fact that something "needs much information to extract from the system to describe it". Suppose an external observer starts with zero information on a system. This is when the observer has maximal uncertainty. When this observer has obtained a certain amount of information, and assuming that the total information of the system plus observer remains constant, the uncertainty on the description of the system has become less, and the change in uncertainty equals minus the information that has been obtained. As complexity of a system equals the information needed to fully describe a system, then complexity equals minus the uncertainty on a system. Defining, in a process of obtaining information from a system, for the information at the end of the process, as $\mathrm{I_f}$, and at the beginning as $I_i$ , one finds that $\mathrm{I_f} - \mathrm{I_i} = -(\mathrm{S_f} - \mathrm{S_i})$ , with $\mathrm{S_f}$ and $\mathrm{S_I}$ the uncertainty (or entropy) of the system at the end of the process and the beginning of the process, respectively. The amount of uncertainty on a system can be quantified in terms of the number of possible outcomes of events arising from a system, with their according probabilities $p_1$, $p_2$, …, $p_n$ . In fact, following Sannon (Shannon 1948), uncertainty on a system, H, is the analogue of entropy, S, and is given by

$$H = S = -K \sum_{i=1}^{n} p_i \, {}^2\log p_i \qquad (1)$$

where $^2\log$ refers to a logarithmic function with base 2, and K is a constant defining the unit of measure conveniently chosen for the problem at hand. Information, and thus complexity, of a system, equals minus H. This was already mentioned in a book by Bar-Yam (Bar-Yam 1997).

For the case of communication between an electronic transmitter and receiver, the case was originally worked out by Shannon (Shannon 1948). For a more thermodynamic/physics background of the concept of information the reader is referred to an earlier publication on the topic by Szilard (Szilard 1929), and also to later publications by Brillouin(Brillouin 1962) and Rothstein (Rothstein 1951). In general terms, following Szilard, we can state that the increase in available information on a complex system equals the decrease of entropy of the system, i.e. the uncertainty on the system. In other words, information on a complex system is the negative of the uncertainty (or entropy) of the system. If one were able to describe a complex system in its entirety, one would have all the information on the system, and have no uncertainty left in fully describing the system. The concepts of entropy, uncertainty, information, and the different possibilities a system can be organized in, are important concepts.

Importantly, information (as is entropy) is an extensive property of a complex system, and is thus scale dependent. At the same time it implies that complexity is also an extensive property of a system, and has a measure that is scale dependent.

## 3. Complex soft matter: stability and transitions

Soft matter considers materials with structural elements interacting with an energy of the order of that of thermal fluctuations. Major progress in the field of soft matter was realized by, amongst others, the Nobel prize winner, De Gennes, who showed that different structural elements can follow the same physics and lead to the same macroscopic behavior (De Gennes 2005). For example, the flow of chocolate and concrete, or the stability of mayonnaise and drilling mud, are governed by the same physics (Sakellariadou 2025).

Complex soft matter deals with materials that are more complex, e.g. containing more different structural elements, with more interactions, or exhibiting a coupling between flow, structure, and moving boundaries. Such complex features make them a convenient starting point to introduce methods utilized in that field, and that will subsequently be discussed to be also applicable to other complex systems.

The complexity in soft matter increases further while considering living matter, like animals and plants, or an ecological system consisting of animals and plants, or the ecology of human beings in cities. Some of the properties of such systems also have been satisfactorily described using the same set of methods.

### 3.1 Micro-emulsions and critical transitions

It is thought illustrative to start with a general sub-class of complex soft matter systems known as “surfactants in solution”. It is an abbreviation for any system that contains surfactants, and one or more liquids. For example, surfactants forming micelles in an aqueous phase, or surfactants in a mixture of water and oil. Like emulsions, but also for example systems that are thermodynamically stable and referred to as microemulsions (Hoar and Schulman 1943, Schulman, Stoeckenius et al. 1959). Such surfactants in solution systems are characterized by the fact that the interactions among its components have a strength in the order of that of thermal energy, with a multitude of different interactions, all of the same order of magnitude. This causes complexity: small changes in conditions result in a different state. Due to the low energies, motion of the components is extensive, leading to an ease for adaptation of the another equilibrium state. Without such extensive motion the system may easily end up in a frozen, strenuous, non-equilibrium state.

For our purpose, we consider a micro-emulsion, firstly described by Hoar and Schulman in 1943, without using the term micro-emulsion itself. The term itself was only given later (Hoar and Schulman 1943, Schulman, Stoeckenius et al. 1959). The microemulsion we consider here is a so called water in oil microemulsion (Zulauf and Eicke 1979) where the droplets are composed of water, and a surfactant layer around them, all embedded in oil. On a smaller scale, monomer surfactants are present in the continuous oil phase, which also contribute to the free energy of the system, but we conclude that these play a minor role in describing the droplet aggregation. This is because the surfactant concentration in the oil phase remains approximately constant over a large range of droplet concentrations. Specifically we will address now the assembly of the micro-emulsion droplets into clusters (Bedaux, Koper et al. 1993, Koper, Sager et al. 1995). The assembly process exhibits a critical concentration below which droplets do not cluster. This critical aggregation on a droplet level is analogous to a critical micelle concentration on a surfactant level. This critical aggregation concentration can be understood in terms of interactions between the droplets. To be more precise, interaction can be quantified by the gain in energy upon joining of two droplets from a situation of two separate droplets. We refrain from a detailed definition of what we mean by energy for our purpose of illustration. More details can be found in the references mentioned. One can distinguish two terms contributing to energy: a so called enthalpy term (favoring joining) and an entropy term (disfavoring joining). Here, the entropy has the same conceptual meaning as in eq.(1). In equilibrium, the energy of a system refers to an organization of its building blocks where energy is minimal, and the enthalpy and entropy are two counteracting elements in the organization of the structural building blocks in the system. The entropy of such a system, and thus its complexity, is determined by the minimization of the energy of the system.

Building further on the example of the micro-emulsion, the aggregates, or clusters, of the droplets, exhibit a droplet concentration dependent size distribution and at a specific, high enough, concentration the cluster with maximum size becomes infinite and results in a so called space spanning cluster. The specific concentration at which such infinite cluster arises is referred to as the *percolation concentration*, referred to as $c_p$, which in fact is a so called critical concentration. It signifies the existence of one (and only one) infinite cluster that is spanning the entire volume of the system. The appearance of such an infinite cluster is related to a transition in macroscopic properties of the system, such as its conductivity, dielectric permittivity, and viscosity, which all exhibit drastic changes upon approaching the percolation concentration. Such properties exhibit scaling near the percolation concentration with $(c-c_p)^{\alpha}$ , where c is the concentration and $\alpha$ a universal exponent (Stauffer 1979). As such, the control of the percolation concentration allows control over the properties, in particular where concentration properties change drastically. We will refer to the case where properties change in the vicinity of a percolation concentration according to $(c-c_p)^{\alpha}$ as critical scaling.

An illustration of the percolation transition in 2 dimensions is depicted in Fig. 1A, with the transition from small clusters to larger clusters, and with the percolation cluster appearing at a critical concentration.

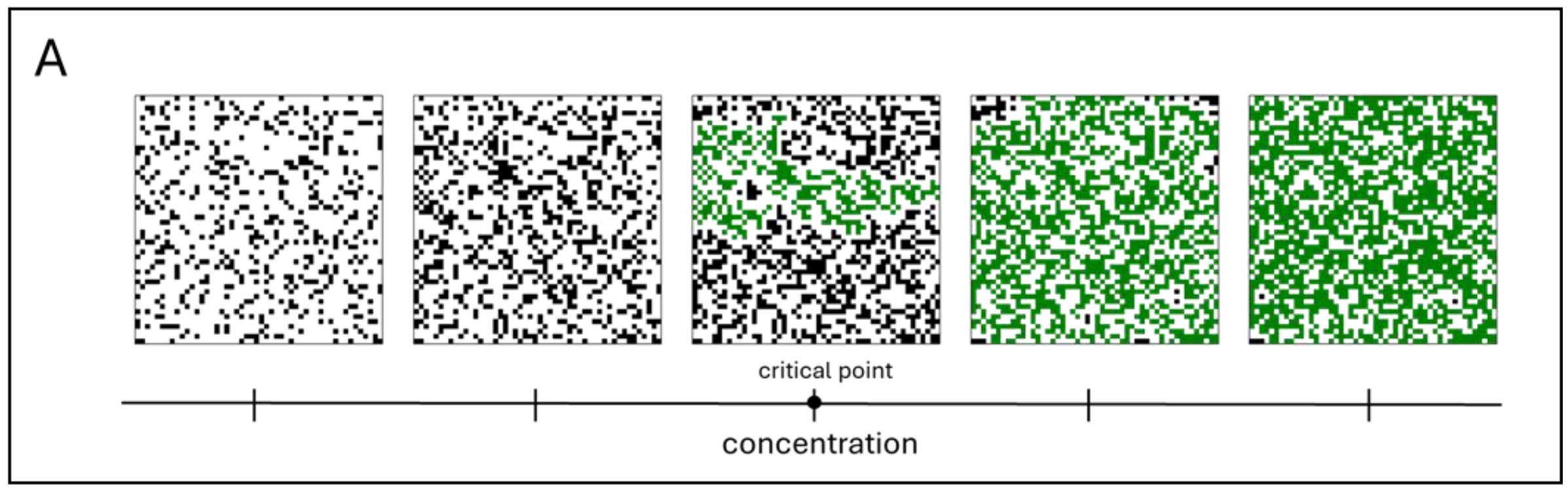

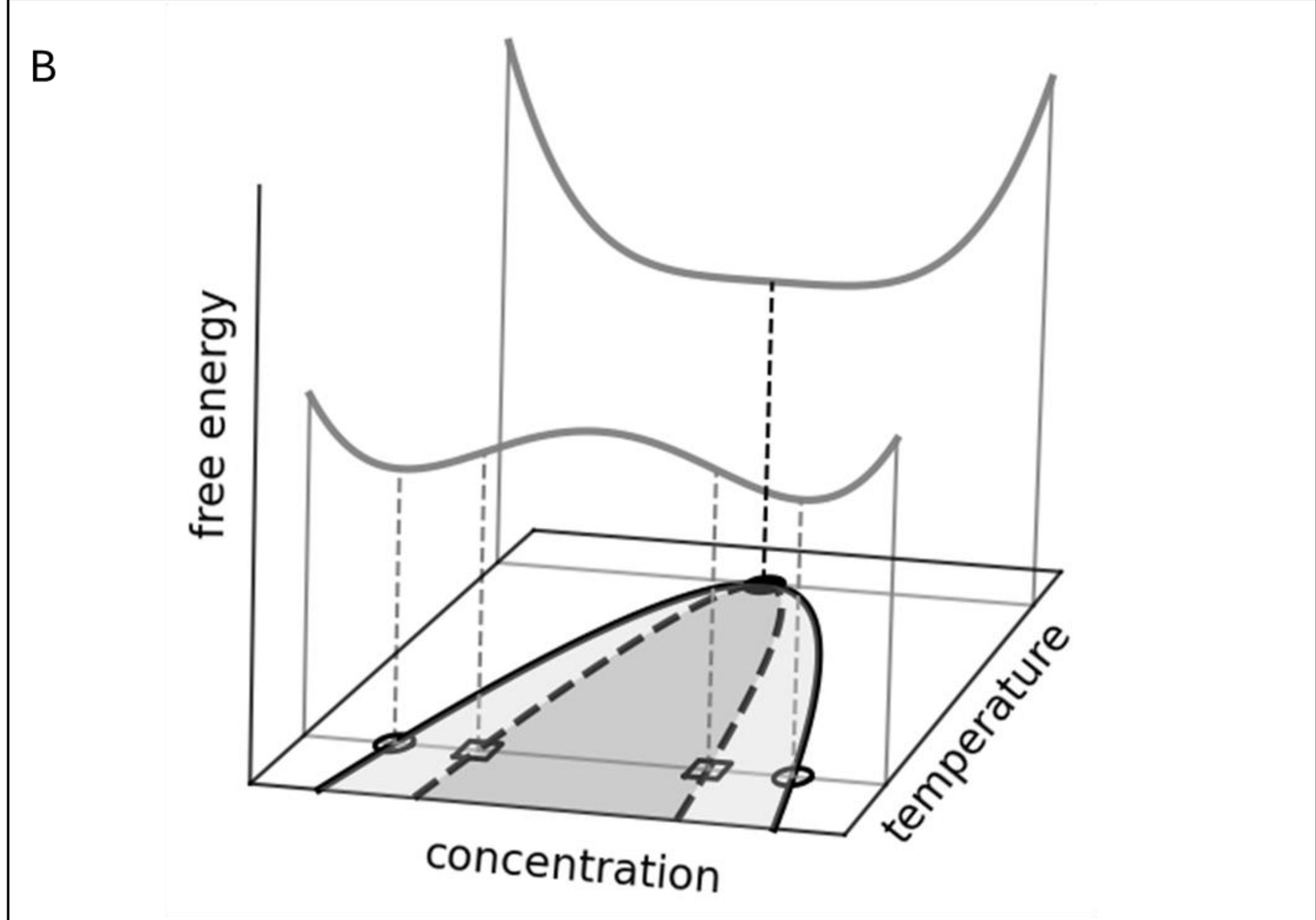

Figure 1. A *a schematic representation of how structures within a system change as a function of increasing concentration of its building blocks (black squares). At a so-called percolation concentration a structure emerges that is connective (percolating) throughout the entire system (green squares). Bcurve representing energy landscape as a function of at specific temperatureThe regions that are separated by the binodal and spinodal lines are the so called stable (white), metastable () and instable () regions. The binodal and spinodal meet in the critical point.*

The clusters are self-similar, i.e. *fractal*, because the mechanism that governs the formation of a cluster, is the same for all n-mers, at any droplet concentrations and for any point in space, at all times. The number of squares filled by a monomer, indicated by $n_m$, in a cluster, scales with the size, R, of the cluster according to $R\sim(n_m)^{D_f}$ where $D_f$ is a number between 1 and 2. In case all squares in a cluster are always be filled, $D_f = 2$. If $D_f < 2$, the cluster becomes more thinly occupied by filled sites the larger the cluster becomes. The value of $D_f$ is referred to as the fractal dimension.

We refer to a more recent work by Cruz et al. (Cruz, Ortiz et al. 2023) that summarizes the values of critical exponents, $\alpha$, and fractal dimensions of clusters, $D_f$, for different types of percolation models. Note that we consider scaling here as a function of concentration of building blocks of the system, and not scaling versus its size. In fact, the occurrence of critical scaling is in the limit of infinite size. In the next section scaling as a function of size is discussed, conveniently applicable to constant concentration of elements in the system. If the elements are organized in a fractal fashion (i.e. forming a structure similar on all scales), a system property is a function of this structure and at the same time, the effect of size of the property is determined by how the structure behaves as a function of size. This in turn is captured by the fractal dimension of the structure formed by the elements.

While approaching the critical point at the critical concentration by lowering the temperature, one observes more and more clusters, and of increasing size. At the critical point, all cluster sizes are present, including one and only one infinite cluster. The according information within the system, in other words the complexity, increases while approaching the critical point. Similarly, one finds that for a liquid-gas transition the critical point exhibits density fluctuations at all wavelengths.

The reason why so-called critical transitions are important is that they are accompanied by drastic (critical) changes in properties of systems. Such transition points signify an instability of the system. Reversely, any drastic change of a property may be signified by a transition in structure. In the case of microemulsions one example of a property drastically changing near the critical percolation transition, is the dielectric permittivity(Koper, Sager et al. 1995).

The next step to consider is what happens to a system where we increase the concentration of the at a temperature lower than the critical temperature. See Figure 1B. At low droplet concentration we will have a stable phase with small clusters and single droplets. When increasing the concentration and following the lower horizontal line in Fig. 1B) we encounter the so-called binodal, and increasing the concentration even further we encounter the spinodal. In the region between the binodal and spinodal, a meta-stable phase exists. Within the spinodal region the system spontaneously demixes into an equilibrium phase with a dilute fraction of droplets, and one concentrated fraction (binodal, Fig 1)). The transition is analogous to a gas-liquid transition. In the latter case it involves molecules, while in the microemulsion case it involves colloidal particles. Such de-mixing behaviour has been observed in other colloidal particle like systems (Verhaegh and Lekkerkerker 1997). Such separations on a molecular and colloidal scale are relevant to understanding the properties of a diverse range of materials, like pharmaceutics, cosmetics, detergents and foods.

The binodal points are defined by having the first derivative of the energy to the concentration being zero (slope of curve of energy versus concentration being zero). This implies that small changes in concentration only increase the energy, i.e. revealing a stable situation. The binodal points have locations where the energy is at a local minimum in the curve.
The spinodal curve is formed by the points that have a second derivative of the energy to the concentration being zero (so called inflexion points). Let us stand still at why this second derivative is another important signature (we note that we return to this point in considering stability conditions in ecological problems). At the inflexion point of the curve, the growth of the slope is halted. After passing the inflexion point, the tendency to return to the minimum becomes less strong. A similar reasoning holds for decreasing concentrations at the second minimum, at the right hand side of the curve. There, an inflexion point exists at the left of the second minimum. In the concentration region between the two inflexion points, the tendency moving towards either the left or the right local minimum becomes stronger upon changing the concentration for a small fraction in that region within the two inflexion points. This means that a small fluctuation in concentration (when situated in between the inflexion points) will automatically lead to the growth of the fluctuation, driving the system towards a separation in two phases, with two local minima at the binodal. The region in between the two inflexion points is referred to as instability region, also known as the spinodal region. The two local minima represent two minimum states of energy for the two separate phases, in equilibrium with one another. In summary, when the second derivative is zero or negative, this signifies a sudden instability transition. The region where the second derivative is positive identifies a stable region.

The principles for this one component system to find the stable, meta-stable and instable regions remain the same for systems containing many different components. The mathematics only becomes more cumbersome as we will outline below.

**3.2. Mixtures with many components and critical transitions.**

Having an arbitrary number of components, thus increasing the complexity of the system, one has booked considerable progress in predicting th critical de-mixing threshold as a function of molecular interactions up to the second virial coefficient, including an exact solution for the spinodal for an arbitrary number of components (Sear and Cuesta 2003, Bot, van der Linden et al. 2024), (Sear and Cuesta 2003, Bot, van der Linden et al. 2024)An important note is that a complex system consisting of many components may not just exhibit an option between one or two phases, but instead multiple phases may exist in equilibrium together. See for example some recent work of Bot et al. (Bot and Venema 2023, Bot, van der Linden et al. 2024). The (phase) transitions from a one phase system towards such multi-phase systems will still show critical scaling but the critical point will extend itself to a line or manifold. These multi-component mixtures are relevant for a range of food material systems that consist of many components, where each component is polydisperse as well. At the same time it bares relevance to understanding the appearance of structures in cells (Carugno, Neri et al. 2022).

For the sake of comparison with analyses performed in ecology, we will write down the main equations for stability transitions in many component molecular mixtures using expressions for contributions to the energy. For multi-component systems, one can use eq. (1) for the expression of the entropy, rewrite $p_i$ in terms of molar concentration for component i as $c_i$, and incorporate mutual interactions between the components i=1, …, N, up to second order in their concentration. One then finds that the free energy, F, for a volume V, can be written as

$$\frac{F}{RTV} = \sum_{i=1}^{N} c_i \cdot ln\, c_i + \sum_{i,j=1}^{N} B_{ij} c_i c_j \tag{2}$$

where the virial coefficients, $B_{ij}$, signify the interaction between the components I and j, with $B_{ij} \equiv B_{ji}$.

This expression is a combination of entropy (cf. first term with eq.(1)), and interactions describing the tendency for components to stick together, or repel one another, in terms of the virial coefficients.

In the case of one component, stability requires that the second derivative of F with respect to concentration be larger than zero. This is the case for all points indicated by a minimum (binodal) in the energy curve in Fig. 1. Instability sets in at concentrations equal, or larger than that, at the so called inflexion points (spinodal), where the second derivative is zero, or smaller than zero, respectively, in Fig. 1. In the complicated case of N components, F becomes a matrix in terms of concentrations of each of the components. The stability criterium in this complicated case becomes that the eigenvalues of the matrix **1**+**B** should all be larger than zero, where the matrix **1** refers to the unity matrix, and **B** refers to the matrix elements of the $i^{th}$ row and $j^{th}$ column, i.e. the virial coefficients $B_{ij}$ (Sear and Cuesta 2003). In other words, the stability criterium becomes an eigenvalue problem. We will come back to the scheme of identifying stability regions when we discuss stability of ecological complex systems, where a description in terms of time derivatives is often used, ut that also resort to a matrix based eigenvalue problem.

A notable and interesting element in stability analyses of many component systems is the use of the so-called Random Matrix theory (RMT) to address eigenvalue problems. This was applied to the co-existence of many components in a liquid by Sear and Cuesta [(Sear and Cuesta 2003). The results of the stability region for a system containing such large number of components can be described in terms of the peak value and width. This was compared to exact expressions for the spinodal as recently determined (Bot, van der Linden et al. 2024). Another noteworthy article appeared by Thewes et al., that discusses how RMT can facilitate understanding the role of a dominantly present component (Thewes, Krüger et al. 2023). Such dominance of interactions was further investigated by Carugno et al., addressing a mixture with interactions among components that are partially structured and partially random (Carugno, Neri et al. 2022). As in the work by Thewes et al., these authors use a categorization of proteins in terms of a family that is of an acidic nature and one that is of a basic nature. They also mention the possibility to consider other main so called families, like being based on hydrophobicity or charge. These considerations are all relevant to many different applications, ranging from phase behaviour in living cells to many component mixtures existing in for example food material systems. Also the application of RMT, in combination with the distinguishing into families of compounds allowing a structuredness as well a randomness to the overall system, is applicable to food material systems as these consist of many components, where each component in addition is polydisperse. Interestingly, following RMT the number of components determines the number of possible phases, but this number is found to be lower than predicted by the Gibbs phase rule, likely because the random interactions used in the RMT effectively introduce competing networks and thus limit the maximal number of phases (Shrinivas and Brenner 2021). It is clear that the role of dominant elements in RMT is currently an active field of research (Carugno, Neri et al. 2022, Thewes, Krüger et al. 2023). An important aspect of recent work on instabilities described by random matrices (Carugno, Neri et al. 2022, Thewes, Krüger et al. 2023) is that full randomness yielded a divergency in the spinodal concentration with the square root of the number of components while incorporating specific interactions yielded a finite critical density of instability.

In the case of dynamics of structural changes and reordering is fast enough, the state will effectively refer to an equilibrium situation, where the averaging over time occurs over all possible states. In such instances, increasing concentration will follow a single phase until the point of separation. In other words, given enough dynamics, the increasing concentration and cluster formation will not be the reason for leading to two separate phases, of which one contains many particles and one only few particles. The reason is that the energy of the system is less when the system contains two phases instead of one. The point at which this transition occurs is the critical point.
If the dynamics is low, a non-equilibrium situation may arise, though possibly still with a spatially percolating structure. However, in such non-equilibrium case, the properties do not necessarily follow critical scaling as the structure may not be indicative of a critical transition point but simply be a sign of a past transition in the spinodal regime (Poulin, Bibette et al. 1999), where a quench of temperature past the critical point is conducted. The knowledge of the equilibrium critical percolation point in systems can be still a guide as to what to expect when a system is quenched into a non-equilibrium state. For example, with applications to food systems among other systems in mind, Loren et al. (Lorén,

Altskär et al. 2001) have convincingly demonstrated that the starting point of the quench, and the path through the phase diagram, dictates the final structure that is found after the quench. In this manner, they could establish which quench paths would lead to bi-continuous versus dispersed structures.

Practically, the above implies that once starting from an equilibrium phase diagram of many components in solution one can end up with a variety of phases on the basis of 1) changing a parameter in an infinitesimal slow manner (equilibrium pathway through the phase diagram) and 2) changing a parameter rapidly, quickly traversing through the phase diagram, where the rate of change determines the final non-equilibrium state. For example like the case of bi-continuous structures that initially arise upon spinodal decomposition, and subsequently quenched into a gel state upon the appropriate temperature quench depth, leading to a gelled bi-continuous structure, with one or both phases being a gel. (Lorén, Altskär et al. 2001). Quenching the trajectory through a phase diagram is clearly applicable to complex food material systems and is often used in practice to arrive at a desired structure and according functionality. For example to arrive at a controlled taste release from gel like materials (van den Berg, van Vliet et al. 2007, Sala, Stieger et al. 2010)

**3.3 Critical slowing down**

Near the percolation threshold, the relaxation behavior of a system upon an external stimulus shows slowing down. In critical material transitions this is caused by the fact that the average cluster sizes increases. In the example case of a microemulsion, droplets form more and larger clusters upon approaching the critical point. Or, in the example of a gas-liquid transition, fluctuations reach larger and larger wavelengths, with according longer and longer relaxation times. Theory predicts how the relaxation times depend on the distance to the critical point (Stauffer 1979, Axelos and Kolb 1990, Stauffer, Coniglio et al. 2005). This occurrence of longer and longer relaxation times was e.g. confirmed in the afore mentioned microemulsion system (Koper, Sager et al. 1995), until very close to the percolation concentration. Close to the percolation concentration, interfacial energy contributions to the overall system energy start to play a role as well, presumably affecting the presence of droplets as the only structure, and allowing other topologies of the interface to emerge.

By way of another example, food gelatin gels have been found to exhibit critical slowing down in terms of changing their elasticity according to a power law with the distance to the critical point determined by a critical temperature(Normand and Parker 2003).

We conclude that the concept of a critical transition induces an experimentally accessible critical slowing down of relaxation (output) signals from a system. As such, time auto-correlation functions can indicate the vicinity of a system to a transition point, even without measuring other properties.

In summary of this section on complex soft matter, we can conclude that the concepts of energy and concentration, as a function of temperature, and the minimization of the energy, built from the concepts of enthalpy and entropy (complexity), play a dominant role in understanding complex soft matter systems, in particular their stability and transitions. Such concepts are considered in equilibrium, whereas non-equilibrium behavior with according microstructural control can be induced from rapidly changing variables, traversing through the equilibrium phase diagram in a non-equilibrium fashion. The concept of a critical transition induces an experimentally accessible critical slowing down of relaxation (output) signals from a system. Several concrete examples of food material systems have been discussed above throughout the text that are food stuffs.

## 4. Living matter, sensory perception and properties of cities: stability and transitions.

Apart from identifying when a system will show a stability instability transition, another proven fruitful way is extracting information on different scales and analyzing the amount of information as a function of scale.

**4.1 Living matter**

So-called scaling analysis has been applied to animal and plant systems, with according scholarly written articles, by West and colleagues (West, Brown et al. 1999, West, Brown et al. 2001). These authors derived information on the internal structures of animals from such scaling analysis, where different types of animals turned out to be scaled versions of themselves in the sense that their metabolic rate scales with a power law of their mass. Notably, in this case of analyzing the scaling of metabolic rate versus mass, the data are retrieved from a system not in equilibrium, but in a quasi-static state, ignoring dynamics on a smaller scale, which however does not affect the understanding on several larger scale properties of the animal.

The scaling analysis as applied to understanding the metabolic rate, MBR, of warm-blooded animals one had found that the MBR was known to scale with their mass, M, according to $MBR \sim M^{3/4}$ (Kleiber 1932). The scaling was later explained in several different ways by West et al. (West, Brown et al. 1999, West, Brown et al. 2001). It can be understood from a principle of minimized effort that balances intake and distribution of energy with its dissipation. It leads to understanding the origin of the self-similar character of the internal vascular structure in warm-blooded animals. Analogously, similar work exists for plants (West, Brown et al. 1999). The scaling even applies to ribosomes in cells (West, Brown et al. 1999). Also, the design of a reindeer nose follows an optimal structural design, for which the analysis explicitly treats entropy production (Magnanelli, Wilhelmsen et al. 2017). In all the afore-mentioned cases, the minimized effort can be related to the unit of energy, the Joule. Furthermore, the minimized effort can, in principle, be related to the interactions between the elementary building blocks of the system. In those cases, one usually lends information from experimental data to reduce the number of possible minimization steps. The experimental data may include scaling laws. Alternatively, a theoretical approach may be possible as well, like West has elegantly shown, that one can arrive at the structural understanding of a complex system like a warm-blooded animal by solely applying scaling arguments, theoretically (West, Brown et al. 1999).

Either way, one utilizes the concept of energy and minimization of effort with the overarching unit of energy. Like in the soft matter cases. It is noted that the minimization of effort is in line with a self-similar structure over which energy or information is transported. Fractality, or self-similarity, facilitates a smooth coupling between different scales, thus avoiding spatially inhomogeneous sinks or sources in entropy production, enabling the system to follow an “energy highway” (Johannessen and Kjelstrup 2005). When the fractality above a certain scale ceases to exist, it signals inhomogeneities above that scale (in terms of properties, functions and concentrations). Interestingly, a transport process such as diffusion for uptake of oxygen in a warm-blooded animal is a smooth process with least effort, initiated by a concentration gradient and the fact that the system has a finite temperature, allowing movement of molecules in the direction of the concentration gradient as understood in terms of statistical mechanics. The process is slow, of course, and any intervention to have it occur faster will generate more entropy and be less effective.

The word effective is important here. Any organization towards least effort is coupled towards the functions that are necessary for that system to execute. Execution will require energy influx and the only question that is important is whether or not a system is able to execute all functions it needs to maintain for its existence, with a minimization of efforts over the entire range of functions that are to be executed for survival.

In order to better understand the behavior of a category of systems it often is a good step to analyze these systems for their input-output versus their size, or versus the scale of the system, if one wishes. In other words, to investigate scaling relations for a range of systems or within one system as a function of an internal variable of the system. Like in the example above of metabolic rate of animals. Usually, if building blocks have been structured according to minimize effort, and this way of structuring has been possible for a range of sizes, the structure will be self-similar over this range of sizes, and a scaling law will be observed for the input-output relationship versus size. As such, a scaling law for input-output versus size signifies a self-similar structure responsible for the input-output relation. This procedure is one obvious way to analyze a set of complex systems. We like to note that in the case of metabolic rate of animals, or output parameters of cities as another example (Lobo, Bettencourt et al. , Bettencourt, Lobo et al. 2010, Bettencourt 2013), one deals with a non-equilibrium system, that however remains in a quasi-static state. This quasi-static state apparently can allow to marginalize contributions that relate to the fine details of underlying structure of the system, leaving out energy related phenomena beneath the scale of the structure that is considered in the analysis. In other words, these energy related phenomena are coarse grained. In, yet, other words, once phenomena on short time scales can be separated from those at long time scales one can arrive at a conveniently simpler description by taking into account only the longer time scales and concomitant larger length scales. Importantly, in many cases the systems are open systems, allowing exchange of energy and matter between the environment and themselves. A quasi-static nature (slow dynamics of such exchange) facilitates a description averaging out the interplay between the system and its environment, analogous to averaging out the interplay between larger scales and smaller scales within a system.

In cases where an energy description or derived notions are not directly possible to describe the observables of the system and in fact quantify minimal effort, other routes using scaling methods have shown their value and thus can be considered to also partially unravel complex systems.

### 4.2 Sensory perception

For a first example where an energy measure cannot be directly applied, but for which information from the system can still be retrieved using scaling related analysis, we mention the sensory perception among human individuals while consuming food. In this case the food system comprises of an individual that is consuming food. Scaling the time with the overall chewing time before swallowing, for each individual, the so called Shannon information as defined in eq. (1) revealed one master curve for all individuals with this (individually determined) scaled time (Sturtewagen, van Mil et al. 2024). Reversely, the findings are suggestive of a possible universal trigger for swallowing originating from a decrease of information gain during chewing. The more general use of the Shannon information measure to sensory perception was recently reviewed (Sturtewagen, van Mil et al. 2025). Analogously, this measure of information was applied to music perceived by individuals (Kulkarni, David et al. 2024). Again, without the need to come to a detailed description of energies involved, but instead using another measure, in this case the Shannon information, one can still deduce understanding and useful information from the system.

### 4.3 Cities and socio-economics

For a second example in this class, economy-related output values by a city do not easily lend themselves for a rigorous energy balance analysis. In the example of a city (Bettencourt 2013), the scaling of city infrastructure, and the scaling of economy related variables, both versus the number of citizens, point to a self-similar structure, provided by the city infrastructure, which reveals how information and matter is transported within the city. The structure arises from a balance between 1) minimization of the use of infrastructural materials, and 2) maximization of the interactions among the inhabitants. Both 1) and 2) in the end are related to an economical cost. The balance has been described and related to the scaling exponents of infrastructural input and economy-related output variables (Bettencourt 2013, Batty 2022). The analysis does not go into the detailed level of an energy balance, and nonetheless suffices to better understand important aspects of the functions of the city, and even provides insight to decide on usefulness of specific system interventions (Bettencourt, Lobo et al. 2010). We note that, if alongside the economic cost one would also include environmental or social costs, the emerging infrastructure may turn out to be different and result in a different balance.

From the above two sections (section 4.2 and 4.3) we learn that one does not necessarily need to obtain all information on all scales of the system to explain properties of the system. The information that is lacking on smaller scales may be either non-changing, or changing at a very slow rate, or negligible compared to the properties at a larger scale, or simply not relevant for the output at hand, or, simply intractable. Nonetheless, lacking such information may still enable one to explain the results observed, like in the case of the cities where many different actors on different levels play a role but nonetheless a simple balance of the main drivers on the largest scale (economy of infrastructure as input and economical advantages as output parameter) could be identified (Bettencourt 2013).

Interestingly, although most likely coincidentally, the number of words in a language being used in a specific book also follows a scaling law with a feature similar to that of percolation. Suppose one word occurring most frequently, one finds that the second frequently used word is occurring half the frequency of the most frequently occurring word, the third occurring a third of the most frequent, etc. This inverse proportionality with the distance to the most frequently occurring word is also found for the size distribution of aggregates close to the percolation transition as obtained from simulations (Watanabe 1996, Watanabe 2002) . The distribution of words has been coined the Zipf distribution, with another well-known and equivalent term of Pareto distribution applied in economics to describe wealth distribution and income distribution.

## 5. Ecological systems: stability and transitions

There are interesting analogues between the mathematical descriptions of material systems and ecological systems regarding their transitions and instabilities (see for example Fort et al. (Fort, Scheffer et al. 2010), and the seminal works by May, for example (May 1972) and (May 1974)). One uses the similar terminologies like stability and instability regions, critical point, and inflexion points.

The mathematical procedures in determining the instability of a material system and an ecological system, have a similar mathematical basis (Stone 2018) as we will shortly address. However, there does not exist a one-to-one mapping of an energy description in physics to a dynamical stability analyses in ecology because the energy contributions in the latter are not all known. We note that the difference between an energy and a dynamical description in general is not an essential difference per se, as in classical mechanics minimization of energy is equivalent to a balance of (generalized) forces and time derivative of the (generalized) momenta (Kibble 1973).

So, the discrepancy between a physics-based description for materials and a description for ecological systems is not just an apparent one. Ecological descriptions utilize dynamical equations, which are phenomenological of nature. The equations in ecology do not find a fundamental basis in an underlying framework as used in physics, the latter illustrated for complex soft matter in the earlier section of this article. The ecological descriptions do not have a basis in energy considerations and according minimization procedures. The parameters are not usually directly experimentally easy to change at will (for example for ethical reasons). Nonetheless, the equations do bear the relevant information that is known for the systems from observations on the appropriate scales. It is exactly this relevant information that allows one to analyze such complex systems. Usually, the equations have the form of a differential equation, identifying a rate of change in time by one or more of the species in a system. One still faces possibilities that one can have a sudden transition upon a change of parameters, similar, but not equivalent to phase transitions in the physics framework. One can also identify instability regions for certain subsets of values of the parameters used in the dynamic equations. Such dynamic instabilities are derived from specific values of the parameters or their combinations in the differential equations. In the physics case, one may induce a temperature change of the system under consideration in Fig. 1, and as a result, depending on the direction and the rate of change, the system may move into a metastable state or an instable state (the two green dotted arrows in Fig. 1. In an ecological system, one may resort to first a specific appropriate dynamic equation and then study its stability regions and instability regions. Usually, one considers the change in number, N, of a species, as a function of time, t, in terms of dN/dt, in interaction with other parts of the system and other species.

Let us for example consider several different species in an ecosystem, competing for the same sources. Such competition can be described by a Lokta-Volterra model according to

$$\frac{dN_i}{dt} = r_i . \frac{N_i}{K_i} \left( K_i - \sum_{j=1}^{\infty} \alpha_{ij} N_j \right) \qquad (3)$$

with $dN_i$ the density of species I, $r_i$ its maximum growth rate per capita, $K_i$ the carrying capacity of species I and $\alpha_{ij}$ the interaction (competition) between species i and j. Interestingly, assuming that all growth rates are equal to one and that all carrying capacities are equal, $K_i = K$, one finds (Fort, Scheffer et al. 2010)

$$\frac{dx_i}{dt} = x_i . \left(1 \quad - \sum_{j=1}^{\infty} \alpha_{ij} x_j \right) \qquad (4)$$

with $x_i = N_i/K$.

In the stability region, i.e. equilibrium, $\frac{dx_i}{dt} = 0$ , and one identifies a set of solution values $x^*_i$ according to

$$x_i^* . \left(1 \quad - \sum_{j=1} \alpha_{ij} x_j^* \right) = 0 \qquad (5)$$

We can study the stability of the equilibria by linearizing equation (4) around its equilibria and study the fate of small perturbations (see for instance Edelstein-Keshet, L. 1988. Mathematical models in biology. 1 edition. McGraw-Hill, Inc., New York.).

The non-diagonal elements of the Jacobian J of equation (4) are $(i \neq j)$:

$$J_{i,j} = \frac{\partial f(x)}{\partial x_j} = -x_i \alpha_{i,j} \qquad (5)$$

where f(x) is the right-hand side of equation (4). The diagonal elements are:

$$J_{i,i} = \frac{\partial f(x)}{\partial x_i} = 1 - \sum_{j=1}^{n} \alpha_{ij} x_j - x_i \alpha_{i,i} \qquad (6)$$

We need to find the eigenvalues of the matrix **J** evaluated in each of the equilibria $x_j^*$. If the real part of the dominant eigenvalue is negative the equilibrium is stable. There is much interest in the occurrence of bifurcations, which are critical parameter values where the model behaviour changes qualitatively. More formally, a bifurcation is the appearance of a topologically nonequivalent phase portrait under variation of parameters ). This means that there is no simple mapping (preserving the direction of time) of the trajectories before and after the bifurcation. This often involves a change in the stability of one or more of the equilibria. At the fold (or saddle-node) bifurcation an unstable equilibrium (saddle) merges with a stable equilibrium and disappears. This implies that at the critical parameter value the equilibrium is stable nor unstable and the dominant eigenvalue is zero.

The pitchfork bifurcation is a special kind of a fold bifurcation as at this critical point two stable equilibria merge with an unstable equilibrium. Such special critical point is equivalent to the critical point identified in the section on complex soft matter, in Fig. 1, i.e. the point where the spinodal and binodal curve meet. In that case, the first and second derivative of the energy versus concentration are both zero.

The analysis above in terms of finding a dominant eigenvalue being smaller than zero is an eigenvalue problem in the end. Such an eigenvalue problem is similar in for stability analysis for multi-component mixtures in the complex soft matter case which also amounted to an eigenvalue problem, with the eigenvalues of the matrix **1**+**B** all being larger than zero in order for the system to remain stable. We note that J and B are not related but the point is that the stability analysis in both cases amounts to an eigenvalue problem. Interestingly, In other words, stability analysis of two completely different systems, i.e. soft matter and ecological, simplifies to an eigenvalue problem, also for arbitrary high number of components.

It is illustrative to consider stability analysis of differential equations where many different types of equations with according different characteristics of their according instabilities exist in solution space. One is able to formulate analogues of instability regions in solution space of differential equations with instabilities as described in the complex soft matter section following an energy-concentration argument. Again, the analysis is different as it refers to an analysis of differential equations.

In general one is interested at first in whether an instability in a system occurs. Since that identifies where a system transitions from 1 specific stable system to at least two specific stable states. This can be analyzed via the stability-instability analysis based on eigenvalues of matrices connecting a observable with internal variables. The endpoint of such an analyses it the prediction whether or not the system remains one phase (homogeneous) or becomes a 2 phase system.

It is noted that more than two phases to be stable at the same time is a possibility. Indeed, having a system of water molecules with three types of biopolymers present one may end up with three phases, as is observed experimentally. In principle, in material systems, the phase rule of Gibbs tells us that the number of possible phases, P, is given in terms of the number of components, N, and the number of degrees of freedom, F, that the system is allowed to have by the formula P=N-F+2.

Whether or not a presence of more than 2 stable phases is possible in ecological systems is a question. This may not be the case in a stable phase, but it is conceivable that, if the rates of change within a system occur inhomogeneously, both metastable and instable regions may co-exist (for a (long) while, at least). Such an effect of dynamics on co-existence of different states was argued to exist in certain ecological situations through a dynamic stability analysis (Rietkerk, Bastiaansen et al. 2021). Also others have reported that changing rates of parameter variation may cause different types instabilities (Scheffer, van Nes et al. 2008), (Siteur, Siero et al. 2014).

Interestingly, the mathematics of stability analysis being equivalent in the physics and ecology area on the basis of matrices and determinants was already clear from the initial work by May on using Random Matrix Theory for arguments on number of species and according ecological stability (May 1972). This work using Random Matrix Theory in ecology was more recently reviewed by Allesina and Tang (Allesina and Tang 2015). For the use of Random Matrix theory in complex soft matter please see above in the according section 3.

Even if quantitative scaling analysis of systems cannot be described in terms of a statistical mechanical framework and proximity to a critical transition point, nonetheless it remains possible to have important clues on the system by means of time dependent measurements (time-autocorrelation measurements).

This can be made plausible by the following argument as previously already outlined in the complex soft matter section 3. There, the critical threshold signified a concomitant slowing down of relaxation times of the system, as the appearance of larger and larger clusters (or correlations within the system), and in the end an infinite large cluster, will go hand in hand with a slowing down of the relaxation. Larger structures (or larger scale correlations) lead to slower relaxation.

Indeed, slowing down near a critical transition is also observed for example for ecological systems (Scheffer, Bascompte et al. 2009, Scheffer, Carpenter et al. 2015) or microbiome systems (Scheffer, Bascompte et al. 2009, Dai, Vorselen et al. 2012, Dai, Korolev et al. 2013, Scheffer, Carpenter et al. 2015). For these systems one does not have access to a free energy description to derive aggregation behavior, from the elementary building block properties and their interactions. Neither may one have access to a more descriptive ecological model. But nonetheless, critical slowing down yields important information as to how a system is evolving and the impact of interventions can be monitored by means of its effects on the slowing down. This occurs in general from the time-autocorrelation function of a given signal. The more and more slowing down of response times is related to occurrence of larger and larger "structures" (regions that exhibit spatial correlations). The occurrence of regions with larger and larger spatial correlations changes the properties of a system, even dramatically when these "structures" become space spanning.

Thus, one conceptual way of obtaining information from complex systems is in identifying early signals for an instability transition, by means of analyzing temporal auto-correlations. This has been used to signify instability transitions in, for example, ecological systems (Scheffer, Bascompte et al. 2009), bacterial systems (Dai, Vorselen et al. 2012, Dai, Korolev et al. 2013), and material systems (Das and Green 2019). Using time dependent output signals one can also analyze other measures, like the Hurst index for geophysical records (Mandelbrot 1969).

At the same time, the dynamical equations can be analyzed in terms of instabilities occurring upon specific values of the observables.

## 6. Food systems: stability and transitions

As how to use the concepts addressed in the current article in particular to food systems on different scales we first like to mention material food systems, such as gels. We like to note that the above-described critical scaling is observed in daily life materials like foods, such as for visco-elastic properties of pectin gels (Axelos and Kolb 1990), and elasticity of gelatin gels (van der Linden and Parker 2005), both as a function of the distance to a critical concentration. The concept of critical scaling was used to interpret elasticity data of various types of protein gels (van der Linden and Sagis 2001).
Furthermore, critical points and phase behaviour of multi-component food material systems was analyzed (Sturtewagen, Dewi et al. 2024).
As another example, on the material properties of food systems, we like to mention the usefulness of understanding structural transitions in providing the possibility of engineering structures in phase separating systems. This has far reaching consequences for, for example, the perception of sweetness (Sala, Stieger et al. 2010). When the structure of a food is such that the distribution of tastants is inhomogeneous this enhances greatly the perception, as for example demonstrated for saltiness of bread (Noort, Bult et al. 2010), and fattiness in a gelled matrix (Mosca, Rocha et al. 2012).

By way of another example, applying slow changes within a system, large food protein particles that are density matched exhibit critical-like clustering and according behaviour (Rouwhorst, van Baalen et al. 2021), although strictly speaking these systems are non-equilibrium. As such, the observed critical behaviour is not strictly speaking critical behaviour, but reminiscent of it. Slowly changing structures may still occur due to reordering in this case, but this resembles a non-equilibrium phenomena, which in the case of accompanying densification is referred to as syneresis (Lucey, van Vliet et al. 1997).

All the above examples refer to a soft matter perspective. When, on the other hand, we consider larger scale food systems, where we should think of a production facility of a specific product, a network of different production facilities for different products, or a network of different production facilities connected to a distribution network of the products and retail companies, the soft matter perspective is only one of the perspectives to be considered. While all entities mentioned may still exhibit an organization according to a minimal effort, including usage of energy and economical aspects, at the same time also other perspectives will be important, such as the socio-economic and infrastructural aspects, similarly to the case of efficiency of output of cities versus the efficiency of its infrastructure, in turn both tied to economic aspects (Bettencourt 2013). The main point is that scaling analysis of the networks of the larger food systems can still be simply conducted in the same fashion as in a material science case, or as in the case of cities.

The step from understanding food soft matter systems to complex food systems is of course huge. Here, food is a tradeable commodity – or even common good or human right (Jackson et al., 2021 ) – in a system in which other key elements are also important. These elements are the food environment, boundary conditions and rules, food system actors (e.g. from farmers to consumers, but also investors and policymakers), food system activities (like production, manufacturing, distribution, marketing, consumption, recycling) and their dynamics (hence, divers temporal specificities of these activities), the inputs or resources or services (needed to produce food) and their overall food system outcomes (availability, affordability, etc.). In a simplified way, these seven elements are similar to the seven building blocks of a game (playing field, rules, players, moves, time, pieces and outcomes) (de Vries 2021). Indeed, gamification was mentioned as a way forward in analyzing the details of complex food systems (van Mil, Foegeding et al. 2014). However, also behavior of actors should not be overlooked since they are involved in the structuring and functioning of food systems. This makes playing games on food systems a challenging exercise. Even more, if one considers that food systems are interacting with other systems like health, (bio-)energy, (bio-)materials, digitalization, etc. – for example in urban systems which are described by Bettencourt (Bettencourt 2013), its complexity increases. If one analyses at the larger scale the efficiency of subsystems, one can still discern important rules for balancing acts between economic and infrastructural aspects of the combined system, for example of a city . All analyses take into account the important notion of focusing on the appropriate scale to discern meaningful rules, and not losing sight of the bigger (or biggest possible) picture.

As an example of delving a bit more into details of a food system, and illustrating the complexity on those more detailed levels, we consider food systems actors that are involved in an imaginary food system. Like interacting particles in a soft food matter, they can be considered as interacting agents in a real-life food system. They can form all kinds of configurations. In a linear chain, one may have farmers, processors, retail, and consumers. In a circular economy, one can add recyclers to close the chain. Traditionally, one observes that food system actors are described via a classical individual business model canvas (Osterwalder and Pigneur 2010). But in reality, purely linear chains or circular loops do not exist. All are interconnected at different scales, linking a diversity of resources to a diversity of food and other bio-based products, to a diversity of markets and different consumer groups. This implies that individually operating food system actors face an enormous number of connections. Consequently, either groupings of food system actors or enlarging business allow incorporating many new actors that together face the nearly unlimited number of connections and seeking optimum positioning of their collective business. The understanding of how groupings of enterprises function dates back to the 1980's, for example by Porter (Porter 1998) who studied territorialized clusters. This work was further detailed for the circular economy coupling clusters to product design (Bocken, de Pauw et al. 2016). For the food and bioeconomy systems, extensive case studies have been performed on food system actor configurations. It became apparent that the number of configurations is not unlimited. Repeatedly, different forms of cooperatives (not only in a single chain like dairy, but also over a range of bio-based products) emerged. Out of 39 case studies on valorization of co-products for food and bio-based products, 6 types of bioeconomy business models were distilled (Donner, Gohier et al. 2020). This number is approximately the square root of case studies performed, as observed for various complex adaptive systems (CAS). These were environmental biorefineries, agriparks, biogas-plant production cluster, upscaling entrepreneurship (for example operating in eco-villages with other actors, which also has been described as CAS (de Vries, Donner et al. 2024) and support structures for multiple actors were recognized (Donner, Gohier et al. 2020). They all shared the vision of jointly innovating to increase the value for bioresources, which resulted in a new concept for circular business model innovation (Donner and de Vries 2021). This combines technological with organizational innovations in different circles of controlling business, co-creating value, influencing the wider environment, and facing planetary and societal challenges.

Coming back to the overall perspective on a larger scale, scaling analyses will teach us the origins of efficiency of the food networks versus their structure in conjunction with external factors like, for example, social-economical. Such insights will be essential to provide clues as to how redesign and mitigate current food systems to become fit for the future, in particular for climate change. In relation to the latter, profile changes, in particular temperature and precipitation, are causing clear differences in food crisis exposure. Interestingly, only taking these two variables into account, it has recently been reported that these two variables are sufficient to predict food crisis exposure by people to sufficient accuracy (Strona 2025). They report that "our findings demonstrate that lack of knowledge about the intermediate nodes that link climate to food security should not prevent us from attempting to explain observed patterns or to make future projections based on available information." This is another example of the possibility that intermediate scale information is not necessarily relevant to understand the relevant aspects on a larger scale. Strona does add, importantly, that the findings should not prevent from investigating more thoroughly the intermediate levels.

## 7. **Smartly using artificial intelligence for complex systems**

Interestingly, the analysis by Strona (Strona 2025) on food crisis exposure versus temperature and precipitation was performed in part using Artificial Intelligence (AI) methodologies. We note that the above-mentioned work performed on efficiency of cities was performed using big data, but not actually using AI (Lobo, Bettencourt et al. , Bettencourt, Lobo et al. 2010, Bettencourt 2013). Instead, the data were hypothesized to follow a scaling law and analyzed accordingly.

One may encounter complex systems, where parts follow a scaling law, and other parts do not. Thus, the system as a whole does not exhibit an overall scaling relation. In such cases, one may resort to a hybrid form by combining domain-based modelling, scaling, and in the case of many data, such as in the case of Strona et al., also AI approaches. Interestingly, hybrid forms are being investigated these days, such as for a food relevant material system using physics domain knowledge encoded in a neural network (Meinders, Yang et al. 2024), alluding to more general applications and arbitrary domains. In particular, a hybrid approach is promising where one encodes the neural networks utilized in the AI model with information that will constrain the number of random possibilities in an arbitrary random neural network. The constraints simply take the network not entirely random, and will act as an efficient starting point for the neural network to establish the best prediction for a given training data set. Such encoding is found to be decreasing the number of data required for a given accurateness of a regular AI model.

The encoding may occur via known physical laws, or simple relations based on known physical laws, elements of them, boundary conditions in outcome parameters, scaling relations, or domain knowledge such as obtained from physics or other domains (Meinders, Yang et al. 2024) (biology, economy, ecology, …). The knowledge may  be explicit knowledge, or years-based experience knowledge by for example experts in chocolate making or other products (Allais, Perrot et al. 2007, Perrot, Trelea et al. 2011).

Neural network (NN) algorithms for optimally predicting input-output relations in general minimize a loss function, as a function of all possible weights annotated to all nodes in the neural network using the Statistical Gradient Descent method. The nature of a neural network loss surface is not well understood currently. Interestingly, in this context, random matrix theory (or RMT, already discussed in sections 3 and 5) is applied to understanding loss surfaces. In fact, the second order Taylor expansion of the loss function in terms of the weights of the nodes in a NN can be written as a matrix, the so called Hessian matrix. RMT is explored to better understand features of the Hessian matrix emerging from a NN upon training.  Some recent research on random matrix models and methods being applied to Hessian matrices of neural networks is reviewed by Leopold (Leopold 2023). We also refer to the references therein (Choromanska, LeCun et al. 2015), (Auffinger, Arous et al. 2011), (Pennington and Bahri 2017), (Granziol 2021 ), (Baskerville, Granziol et al. 2021). Leopold concludes that "random matrix theory is a useful tool for various machine learning applications and it is a fruitful field of mathematics to explore, in particular, in the context of theoretical deep learning".

In the context of hybrid NN, and in particular encoded NN, it is worthwhile to explore the characteristics of the Hessian matrix from NN's as a function of the level of (function based) encoding of the NN. In fact, functional encoding of specific neurons (by for example changing one of the usual ReLU functions to an n-th order algebraic function) will affect the second derivative of at least one element of the Hessian matrix of the loss function. Following Pennington and Bahri (Pennington and Bahri 2017), the width of the Machenko-Pastur distribution of eigenvalues of the Hessian will be increasing by this functional encoding. The Machenko-Pastur distribution of eigenvalues of the Hessian can be explained by RMT (section 3 and 5). To calculate whether a specific functional encoding causes an observable deviation from the Machenko-Pastur distribution one can use a paper by Baik et al. (Baik, Ben Arous et al. 2005). If such an observable deviation occurs one refers to the occurrence of a rank 1 spike, effectively exhibiting a dominant feature in the network. Such dominant features and rank 1 spikes may also arise from a network with only ReLU functions, but only a combination of ReLU's is capable of causing such dominant feature. The deviations from the Machenko-Pastur distribution, i.e. spikes, reveal structuredness while the Machenko-Pastur distribution reveals the characteristics linked to randomness. In general one can conclude that the Hessian, and its analysis using RMT, provides a measure for the structure of the NN and for dominant features. It will be interesting to investigate to what extent the weights of the ReLU function neurons surrounding the encoded neuron and the Hessian structure itself will be changed by functional neuron encoding.

The above will shed light to what extent a specific encoding of a NN affects the neural network. And to what extent encoding changes the level of structuredness and randomness in the trained NN. In general, deeper understanding of NN using RMT will be helpful in further understanding how NN's work and to further optimize and design the architecture of a neural network.

Regarding randomness and structuredness in the context of networks, it is interesting to note that a meso-structure could be induced in a NN by means of attributing synaptic weights from a probability distribution to the nodes (Aljadeff, Stern et al. 2015, Aljadeff, Renfrew et al. 2016). So, the operation rules for nodes can determine at least partially a structuredness of the NN.

Structuredness and randomness were also discussed in the section 3, where dominance either in concentration and/or concentration of components, in understanding de-mixing phenomena, was induced by splitting the RM into a structured, non-random matrix and a random matrix. In fact, such distinction between randomness and structuredness of interactions was also addressed previously by Allesina and Tang in the context of predicting ecological stability (section 5) by comparing the consequences of a complete set of random interactions with the case where some interactions are dominant, and gives more structure to the system (Allesina and Tang 2012). Such comparison was more recently addressed by Baron and Galla (Baron and Galla 2020) for ecological systems.

We conclude that the random matrices are a tool to study and predict behaviour of complex systems, ranging from complex soft matter to ecology and NN's.

## 8. Complex systems not close to a critical transition

In cases where the system is not close to a critical transition, one can still investigate the size dependency of input-output of a set of similar systems and scaling of input-response for a system. If a scaling law exists, one still can defer to an underlying structure responsible for a general principle along which the systems (self-)organizes. If neither the systems exhibit critical scaling nor exhibit scaling of input-output or output versus size, the systems are on the large scale not self-similarly structured. Structures may still exist in different sizes, but there does not exist self-similar structures on a larger scale. In other words, the overall structure is inhomogeneous. Or. self-organization to the scale of the observation apparently has not taken place under the circumstances the system has been exposed to. Inhomogeneities form one of the reasons as to why scaling is not observed. If the fraction of inhomogeneities is small, scaling may still be observable, but more approximately. In case of non-scaling, one can understand such systems by resorting to for example effective medium theories, which are based on a mean field approach to describe the action on one element in terms of a mean field formed by the surrounding elements and their embedding matrix. This has been scholarly worked out for electric phenomena by Landauer (Landauer 1978), but the approach is valid generally.

## 9. Summary and conclusions

We conclude that one can follow a systematic approach to explore complex systems. Firstly, one should explore scaling relations between input and responses for a system. For example, the elastic response as a function of changing concentration of particles in the system. If one encounters scaling at some range of scales, self-organization may be at hand under the circumstances the system has been exposed to (equilibrium or non-equilibrium). The scaling may be utilized to explore a) the drivers of the self-organization and b) the structures that determine properties of, and transport within, the system.

Secondly, one can explore scaling relations between inputs and responses within a category of systems of varying size and exhibiting similarities in properties. For example, warm-blooded animals of different species and according different mass exhibit scaling of their metabolic rate versus their mass. The scaling signifies self-organisation in a specific manner that is general over this class of animals.

If scaling is not found, one may resort to smaller (sub-)systems or larger systems to check for scaling. This is analogous to the remark by Anderson on the importance of symmetry breaking in complex systems (Anderson 1972, Anderson 1995). If one does not encounter scaling laws at all, one can conclude that the system does not bare any regular (self-similar) structure and is an assembly of separate pieces or that inhomogeneities are present to such a large extent that scaling is not relevant for the system. In these cases, one may resort to mean field approaches.

Thirdly, scaling between input and responses of a system may exhibit a critical point, signifying vastly different behaviour below or above that point, signifying an instability when crossing from below to above that point. For example, a particle system that exhibits gelation only above a specific concentration of particles. The identification of such a critical point is key to describing the properties of a system. It has been illustrated how the identification of instability regions for soft matter and ecological systems, both containing possibly very high number of components, simplifies to a matrix eigenvalue problem.

If one cannot identify a critical point regarding input versus responses, an indication for its existence can still be drawn from observing critical slowing down, i.e. the divergence of relaxation times of the system. Formally, this amounts to analyze the temporal auto-correlation functions of response signals. This critical slowing down is, for example, illustrated in droplet microemulsion systems around the critical droplet concentration and in ecological systems close to instability.

Obvious requirements for all of the above strategies are extracting quantitative information of the system at different scales (i.e. real data), formulating hypotheses on the reasons of the scaling relations observed, and developing theories, possibly accompanied with simulations, explaining those reasons. It is important to note that the information (and data) that is required to understand a system is dependent on what function of a system is of interest. Not all scales are necessary to start modelling a system, but it depends on what one intends to model.

The above strategies are relevant to addressing complex systems and their big data via hypothesized driven approaches, using scaling arguments. Apart from scaling analysis, in particular when scaling is not detectable, i.e. there does not exist any self-similar structure, one may resort to other approaches to detect some structuredness in a complex system.

One of the underlying mathematics for addressing complex multi-component systems and detecting structuredness follows a well-established framework of RMT, that has recently been extended to account for some dominant features on top of a random behaviour around some distribution. Such approaches do justice to how nature often reveals itself. The framework of RMT also provides a way to analyse the structuredness and randomness in a NN and a way to quantitatively determine the effects of encoding a NN, or of other ways for providing information to a NN.

The above summarized set of methods facilitates to identify their level of structuredness and randomness in complex systems. It helps to better predict upcoming transitions in complex systems, according critical points, and sudden instabilities. As such, it facilitates in extracting information from a system, before, during and after the time that one makes an intervention, which in turn will help to decide which interventions are best to maintain or change functions of a complex system.

Further progress on addressing complex challenges like mitigating climate change in relation to sustainable food production is foreseen by using the universal concepts and combining analytical matrix features of dominance versus randomness, domain specific encoding of neural networks and identifying dominancy in neural networks using the analytical matrix features of dominance versus randomness.

**Acknowledgement**

We acknowledge discussions with profs. Marten Scheffer and Rutger van Santen, and dr. Monique Axelos.